\documentclass[aos]{imsart}

\usepackage{amsthm,amsmath,natbib}
\RequirePackage[colorlinks,citecolor=blue,urlcolor=blue,breaklinks]{hyperref}

\arxiv{}

\RequirePackage{hypernat}

\startlocaldefs
\usepackage{latexsym,amssymb,amsfonts,bbm}
\usepackage{graphicx,psfrag}
\usepackage{graphics,enumerate}

\begin{document}

\begin{frontmatter}

\title{Peter Hall's work on high-dimensional data and classification}
\runtitle{High-dimensional data and classification}
\begin{aug}
\author{\fnms{Richard} J. \snm{Samworth}\thanksref{t2,m1}\ead[label=e1]{r.samworth@statslab.cam.ac.uk}}
\ead[label=u1,url]{http://www.statslab.cam.ac.uk/\~{}rjs57}
\thankstext{t2}{The research of Richard J. Samworth was supported an EPSRC Early Career Fellowship and a Philip Leverhulme prize.}
\runauthor{R. J. Samworth}
\affiliation{University of Cambridge\thanksmark{m1}}

\address{Statistical Laboratory \\ Wilberforce Road \\ Cambridge \\ CB3 0WB \\ United Kingdom\\ 
          \printead{e1}\\ \printead{u1}}

\end{aug}

\begin{abstract}
In this article, I summarise Peter Hall's contributions to high-dimensional data, including their geometric representations and variable selection methods based on ranking.  I also discuss his work on classification problems, concluding with some personal reflections on my own interactions with him.
\end{abstract}


\end{frontmatter}

\section{High-dimensional data}
\label{Sec:HighDim}

Peter Hall wrote many influential works on high-dimensional data, though notably he largely eschewed the notions of sparsity and penalised likelihood that have become so popular in recent years.  Nevertheless, he was interested in variable selection, and wrote several papers that involved ranking variables in some way.  Perhaps his most well-known papers in this area, though, concern geometrical representations of high-dimensional data.

\subsection{Geometric representations of high-dimensional data}

\citet{HallLi1993} was one of the pioneering works in the early days of high-dimensional data analysis that tried to understand the properties of low-dimensional projections of a high-dimensional isotropic random vector $X$ in $\mathbb{R}^p$.  As motivation, let $\gamma \in \mathbb{R}^p$ have $\|\gamma\| = 1$ and suppose that 
\begin{equation}
\label{Eq:Linear}
\forall b \in \mathbb{R}^p, \exists \alpha_b,\beta_b \in \mathbb{R}, \quad \mathbb{E}(b^TX|\gamma^T X = t) = \alpha_b t + \beta_b.
\end{equation}
This condition says that the regression function of $b^TX$ on $\gamma^TX$ is linear.  Then, using the isotropy of $X$, 
\[
0 = \mathbb{E}(b^TX) = \mathbb{E}\{\mathbb{E}(b^TX|\gamma^TX)\} = \mathbb{E}(\alpha_b\gamma^TX + \beta_b) = \beta_b.
\]
Moreover,
\[
b^T\gamma = \mathrm{Cov}(b^TX,\gamma^TX) = \mathbb{E}\{\mathbb{E}(b^TXX^T\gamma|\gamma^TX)\} = \alpha_b \gamma^T\mathbb{E}(XX^T)\gamma = \alpha_b,
\]
and we conclude that $\mathbb{E}(X|\gamma^TX=t) = t\gamma$, or equivalently, 
\begin{equation}
\label{Eq:Norm}
\|\mathbb{E}(X|\gamma^TX=t)\|^2 - t^2 = 0.
\end{equation}
The left-hand side of~\eqref{Eq:Norm} is always non-negative, so can be used as a measure of the extent to which the condition~\eqref{Eq:Linear} holds.  Remarkably, under very mild conditions on the distribution of $X$, \citet{HallLi1993} proved that if $\gamma$ is drawn from the uniform distribution on the unit Euclidean sphere in $\mathbb{R}^p$, then
\[
\|\mathbb{E}(X|\gamma, \gamma^TX=t)\|^2 - t^2 \stackrel{p}{\rightarrow} 0
\]
as $p \rightarrow \infty$.  This is equivalent to the statement
\[
\sup_{b \in \mathbb{R}^p:\|b\|=1,b^T\gamma=0} \bigl|\mathbb{E}(b^TX|\gamma, \gamma^TX=t)\bigr| \stackrel{p}{\rightarrow} 0
\]
as $p \rightarrow \infty$.  See also \citet{DiaconisFreedman1984}, who showed that under mild conditions, most low-dimensional projections of high-dimensional data are nearly normal.  Of course, when $X$ has a spherically symmetric distribution,~\eqref{Eq:Linear} holds for every $\gamma \in \mathbb{R}^p$ with $\|\gamma\| = 1$.  But the result of this paper shows that even without spherical symmetry, there is a good chance (in the sense of random draws of $\gamma$ as described above) that~\eqref{Eq:Linear} holds, at least approximately, when $p$ is large.  An important statistical consequence of this is that even if the relationship between a response $Y$ and a high-dimensional predictor is non-linear, say $Y = g(\gamma^TX,\epsilon)$ for some unknown link function $g$ and error $\epsilon$, standard linear regression procedures can often be expected to yield an approximately correct estimate of $\gamma$ up to a constant of proportionality.  The generalisation of this result that replaces $\gamma^TX$ with $\Gamma^TX$, where $\Gamma$ is a random $p \times k$ matrix with orthonormal columns, also plays an important role in justifying the use of sliced inverse regression for dimension reduction \citep{Li1991}. 

Another seminal paper that articulated many of the key geometrical properties of high-dimensional data is \citet{HMN2005}.  This paper begins with the simple, yet remarkable, observation that if $Z \sim N_p(0,I)$, then $\|Z\| = p^{1/2} + O_p(1)$ as $p \rightarrow \infty$.  Thus, data drawn from this distribution tend to lie near the boundary of a large ball.  Similarly, the pairwise distances between points are almost a deterministic distance apart, and the observations tend to be almost orthogonal.  In fact, the authors go on to explain that, under much weaker assumptions than Gaussianity, the data lie approximately on the vertices of a regular simplex, and that the stochasticity in the data essentially appears as a random rotation of this simplex.  As well as clarifying the relationship between Support Vector Machines \citep[e.g.][]{ChristianiniST2000} and Distance Weighted Discrimination classifiers \citep{MTA2007} in high dimensions, the paper forced researchers to rewire their intuition about high-dimensional data, and precipitated a flood of subsequent papers on high-dimensional asymptotics.

\subsection{Variable selection and ranking}

The last 15 years or so have seen variable selection emerge as one of the most prominently-studied topics in Statistics.  Although Peter's instinct was to think nonparametrically, he realised that he could contribute to a prominent line of research in the variable selection literature, namely marginal screening \citep[e.g.][]{FanLv2008,FSW2009,LZZ2012}, via the deep understanding he developed for rankings.  \citet{HallMiller2009a} defined variable rankings through their generalised correlation with a response, while \citet{DelaigleHall2012} studied variable transformations prior to ranking based on correlation as a method for dealing with heavy-tailed data.  For classification, \citet{HTX2009a} proposed a cross-validation based criterion for assessing variable importance, while in the unsupervised setting, \citet{ChanHall2010} suggested ranking the importance of variables for clustering based on nonparametric tests of modality.  

These works above were underpinned by Peter's realisation that he could explain how perhaps his favourite tool of all, namely the bootstrap, could be used to quantify the authority of a ranking \citep{HallMiller2009b}.  In fact, there are some subtle issues here, particularly surrounding the issue of ties.  Peter developed an ingenious method for proving that even though the standard $n$-out-of-$n$ bootstrap does not handle this issue well, the $m$-out-of-$n$ bootstrap overcomes it in an elegant way.

\section{Classification problems}

I believe that Peter may have become interested in classification problems in the early 2000s at least partly through ideas of bootstrap aggregating, or bagging \citep{Breiman1996}.  Indeed, in \citet{FriedmanHall2007}, a preprint of which was already available in early 2000, Peter had attempted to understand the effect of bagging in $M$-estimation problems.  This is a typical example of Peter's extraordinary ability to explain empirically observed effects through asymptotic expansions.  One of the other interesting contributions of this work is that subsampling (i.e.\ sampling without replacement) half of the observations closely mimics ordinary $n$-out-of-$n$ bootstrap sampling, a very useful fact that has been observed and exploited in several other contexts, including stability selection for choosing variables in high-dimensional inference \citep{MeinshausenBuhlmann2010,ShahSamworth2013} and stochastic search methods for semiparametric regression \citep{DSS2013}. 

Classification problems are ideally suited to bagging, because the discrete nature of the response variable means that small changes to the training data can often yield different outputs from a classifier; in the terminology of \citet{Breiman1996}, many classifiers are `unstable'.  Suppose we are given training data $\mathcal{X} := \{(X_1,Y_1),\ldots,(X_n,Y_n)\}$, where each $X_i$ is a covariate taking values in a general normed space $\mathcal{B}$, and $Y_i$ is a response taking values in $\{-1,1\}$.  Assume further that we have access to a classifier $\hat{C}_n(\cdot) = \hat{C}_n(\cdot;\mathcal{X})$ constructed from the training data, so that $x \in \mathcal{B}$ is assigned to class $\hat{C}_n(x;\mathcal{X})$.  To form the bagged version $\hat{C}_n^*$ of the classifier, we draw $B$ bootstrap resamples $\{\mathcal{X}_b^*:b = 1,\ldots,B\}$ from $\mathcal{X}$, and set
\[
\hat{C}_n^*(x) := \mathrm{sgn}\biggl(\frac{1}{B}\sum_{b=1}^B \hat{C}_n(x;\mathcal{X}_b^*)\biggr).
\]
Peter got me interested in bagging nearest neighbour classifiers.  Ironically, the nearest neighbour classifier had been described by Breiman as stable, since the nearest neighbour appears in more than half --- in fact, around $1 - (1-1/n)^n \approx 1 - e^{-1}$ --- of the bootstrap resamples; thus the bagged nearest neighbour classifier is typically identical to the unbagged version.  In \citet{HallSamworth2005}, however, we studied the effect of drawing resamples (either with or without replacement) of smaller size $m$.  Naturally, this reduces the probability of including the nearest neighbour in the resample, and the bagged classifier is now well approximated by a weighted nearest neighbour classifier with geometrically decaying weights; see also \citet{BiauDevroye2010}.  In order for bagging to yield any asymptotic improvement over the basic nearest neighbour classifier, we require $m/n < 1/2$ (when sampling without replacement) and $m/n < \log 2$ (when sampling with replacement); in order to converge to the theoretically-optimal Bayes classifier, we require $m = m_n \rightarrow \infty$ but $m/n \rightarrow 0$. 

Once classification problems had piqued his interest, Peter set about trying to answer some of the key questions on rates of convergence and tuning parameter selection that would naturally have occurred to him given his earlier work on nonparametric inference.  \citet{HallKang2005} studied the performance of classifiers constructed from kernel density estimates of the class conditional distributions on $\mathcal{B} = \mathbb{R}^d$.  A particularly curious discovery he made there is that even in the simplest case where $d=1$ and where the class conditional densities $f$ and $g$ cross only at the single point $x_0$, the rate of convergence and order of the asymptotically optimal bandwidth depends on the sign of $f''(x_0)g''(x_0)$.  In \citet{HPS2008}, we considered similar problems in the context of $k$-nearest neighbour classification, obtaining an asymptotic expansion for the regret (i.e.\ the difference between the risk of the $k$-nearest neighbour classifier and that of the Bayes classifier) which implied that the usual nonparametric error rate of order $n^{-4/(d+4)}$ was attainable with $k$ chosen to be of order $n^{4/(d+4)}$.  The form of the expansion made me realise that the limiting ratio of the regrets of the bagged nearest neighbour classifier and the $k$-nearest neighbour classifier (with both the resample size $m$ and the number of neighbours $k$ chosen optimally) depended only on $d$, and not on the underlying distributions.  To my great surprise, this limiting ratio was greater than 1 when $d=1$, equal to 1 when $d=2$ and less than 1 for $d \geq 3$ (though approaching 1 for large $d$).  It took me some years to explain this phenomenon in terms of the optimal weighting scheme \citep{Samworth2012}.

In more recent years, Peter turned his attention to a wealth of other important, though perhaps less well studied, issues in classification.  Some of these were motivated by what he saw as drawbacks of existing classifiers.  For instance, in \citet{HTX2009b}, he developed classifiers based on componentwise medians, to alleviate the difficulties of both computing and interpreting multivariate medians; such methods can be highly effective for high-dimensional data that may have heavy tails.  In \citet{ChanHall2009a}, he studied robust versions of nearest neighbour classifiers for high-dimensional data that try to perform an initial variable selection step to reduce variability.  \citet{ChanHall2009b} presented simple scale adjustments to make distance-based classifiers (primarily designed to detect location differences) less sensitive to scale variation between populations; see also \citet{HallPham2010}.  \citet{HallXue2010} and \citet{HXX2013} concerned settings where one might want to incorporate the prior probabilities into a classifier, and where these prior probabilities may be significantly different from $1/2$, respectively.  Finally, \citet{GhoshHall2008} discovered the phenomenon that estimating the risk of a classifier, and estimating the tuning parameters to minimise that risk, are two rather different problems, requiring the use of different methodologies. 

\section{Some personal reflections}

I first met Peter as a PhD student when he visited Cambridge in 2002.  I spent an hour or so discussing a problem I was working on that involved using ideas of James--Stein estimation to find small confidence sets for the location parameter of a spherically symmetric distribution \citep{Samworth2005}.  I was blown away at the speed with which he was able to understand where my difficulties lay, and make helpful suggestions.  Shortly afterwards, he invited me to spend six weeks at the Australian National University in Canberra in July--August 2003.  I arrived utterly exhausted after nearly 24 hours in the air, but Peter was full of energy when he kindly picked me up from the bus station.  Almost the first thing he said to me was: `I've got a problem I thought we could think about...', and he proceeded to take out a pen and pad of paper; one couldn't help but be drawn along by his enthusiasm for research.

Everything with Peter happened at breakneck speed, whether it was dashing around the supermarket, a driving tour through the rural Australian Capital Territory or, of course, writing papers.  Many of his collaborators will have experienced discussing a problem with Peter one evening and returning to the office the following morning to find that he had typed up a draft manuscript that would form the basis of a joint paper.  His prose was always elegant, and he had a wonderful ability to see his way through technical asymptotic arguments, aided by almost physicist-like intuition for what ought to be true.

\begin{figure}[ht!]
\begin{center}
\includegraphics[width=\textwidth]{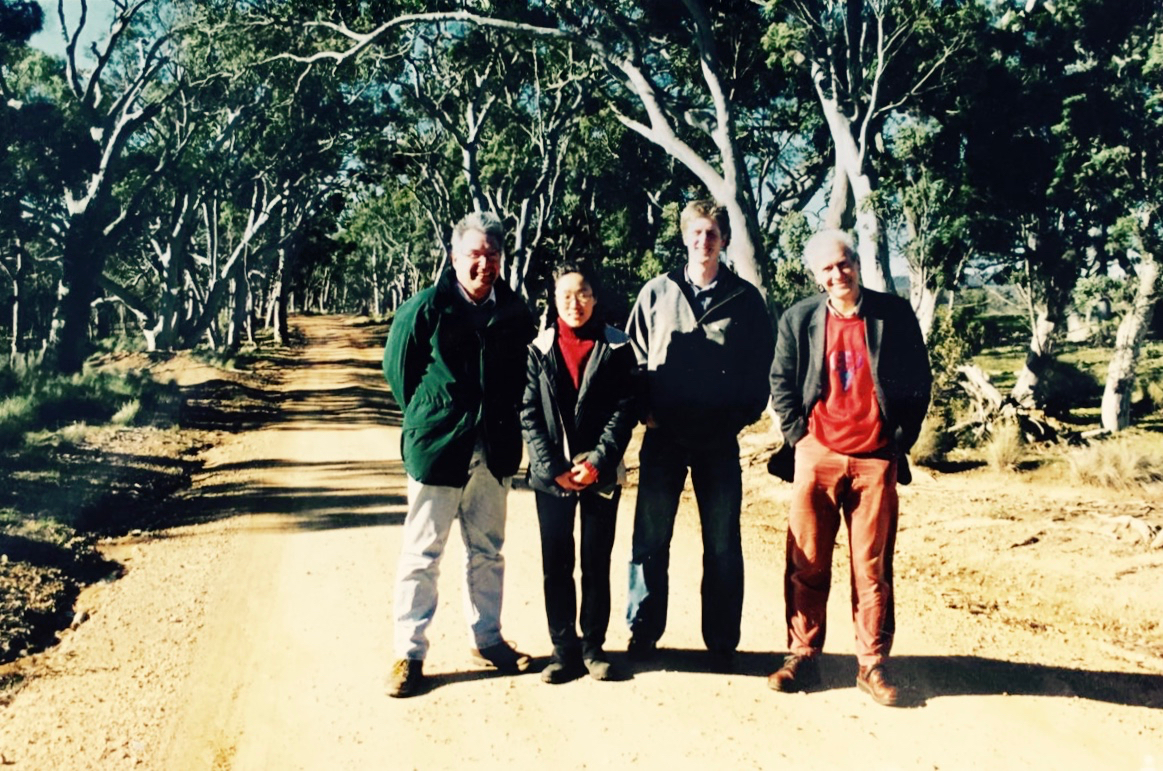}
\end{center}
\caption{Peter with Juhyun Park (Lancaster University), the author and Nick Bingham (Imperial College London) on a blustery day in rural Australian Capital Territory in 2003.}
\end{figure}

One of my favourite Peter stories, which I initially heard second-hand but which he later confirmed was true, concerned a time when he'd been asked to teach an elementary Statistics course to students with really very little quantitative background.  Realising that he'd lost some of the students along the way, and in order not to ruin their grades, Peter had a cunning idea and spent the last class before the final going through the problems that he'd set on the exam.  To his horror, however, the students still flunked the exam.  When Peter bumped into one of the students and asked in bemusement `What happened?  I went through the questions in the last class', the student replied `Yes, but you did them in a different order'!

Peter had seemingly boundless energy and capacity to work, but he was also a very gentle individual in many ways.  He was extraordinarily generous to others, and particularly junior researchers for whom he did so much.  He was a remarkable person and I miss him very deeply.


\begin{thebibliography}{99}
\bibitem[{Biau and Devroye(2010)}]{BiauDevroye2010}Biau, G. and Devroye, L. (2010) On the layered nearest neighbour estimate, the bagged nearest neighbour estimate and the random forest method in regression and classification
\newblock \emph{J. Mult. Anal.}, \textbf{101}, 2499--2518.

\bibitem[{Breiman(1996)}]{Breiman1996}Breiman, L. (1996) Bagging predictors.
\newblock \emph{Mach. Learn.}, \textbf{24}, 123--140.

\bibitem[{Chan and Hall(2009a)}]{ChanHall2009a}Chan, Y.-b. and Hall, P. (2009a) Robust nearest-neighbor methods for classifying high-dimensional data. 
\newblock \emph{Ann. Statist.}, \textbf{37}, 3186--3203.

\bibitem[{Chan and Hall(2009b)}]{ChanHall2009b}Chan, Y.-B. and Hall, P. (2009b) Scale adjustments for classifiers in high-dimensional, low sample size settings. 
\newblock \emph{Biometrika}, \textbf{96}, 469--478.

\bibitem[{Chan and Hall(2010)}]{ChanHall2010}Chan, Y.-b. and Hall, P. (2010) Using evidence of mixed populations to select variables for clustering very high dimensional data. 
\newblock {J. Amer. Statist. Assoc.}, \textbf{105}, 798--809.

\bibitem[{Christianini and Shawe-Taylor(2000)}]{ChristianiniST2000}Christianini, N. and Shawe-Taylor, J. (2000) \emph{An Introduction to Support Vector Machines}.
\newblock Cambridge University Press, Cambridge.

\bibitem[{Delaigle and Hall(2012)}]{DelaigleHall2012}Delaigle, A. and Hall, P. (2012) Effect of heavy-tails on ultra high dimensional variable ranking methods. 
\newblock \emph{Statistica Sinica}, \textbf{22}, 909--932.

\bibitem[{Diaconis and Freedman(1984)}]{DiaconisFreedman1984}Diaconis, P. and Freedman, D. (1984) Asymptotics of graphical projection pursuit.
\newblock \emph{Ann. Statist.}, \textbf{12}, 793--815.


\bibitem[{D\"umbgen, Samworth and Schuhmacher(2013)}]{DSS2013}D\"umbgen, L., Samworth, R. J. and Schuhmacher, D. (2013) Stochastic search for semiparametric linear regression models. 
\newblock In \emph{From Probability to Statistics and Back: High-Dimensional Models and Processes -- A Festschrift in Honor of Jon A. Wellner}. Eds M. Banerjee, F. Bunea, J. Huang, V. Koltchinskii, M. H. Maathuis, pp. 78--90.

\bibitem[{Ghosh and Hall(2008)}]{GhoshHall2008}Ghosh, A. K. and Hall, P. (2008) On error-rate estimation in nonparametric classification. 
\newblock{Statistica Sinica}, \textbf{18}, 1081--1100.

\bibitem[{Fan and Lv(2008)}]{FanLv2008}Fan, J. and Lv, J. (2008) Sure independence screening for ultrahigh dimensional feature space (with discussion).
\newblock \emph{J. Roy. Statist. Soc. Ser. B}, \textbf{70}, 849--911.

\bibitem[{Fan, Samworth and Wu(2009)}]{FSW2009}Fan, J., Samworth, R. and Wu, Y. (2009) Ultrahigh dimensional feature selection: beyond the linear model.
\newblock \emph{J. Mach. Learn. Res.}, \textbf{10}, 2013--2038.

\bibitem[{Friedman and Hall(2007)}]{FriedmanHall2007}Friedman, J. H. and Hall, P. (2007) On bagging and nonlinear estimation. 
\newblock {J. Statist. Plann. Inf.}, \textbf{137}, 669--683.

\bibitem[{Hall and Kang(2005)}]{HallKang2005}Hall, P. and Kang, K.-H. (2005) Bandwidth choice for nonparametric classification. 
\newblock \emph{Ann. Statist.}, \textbf{33}, 284--306.

\bibitem[{Hall and Li(1993)}]{HallLi1993}Hall, P. and Li, K.-C. (1993) On almost linearity of low dimensional projections from high dimensional data.
\newblock \emph{Ann. Statist.}, \textbf{21}, 867--889.

\bibitem[{Hall, Marron and Neeman(2005)}]{HMN2005}Hall, P., Marron, J. S. and Neeman, A. (2005) Geometric representation of high dimension, low sample size data.
\newblock \emph{J. Roy. Statist. Soc. Ser. B}, \textbf{67}, 427--444.

\bibitem[{Hall and Miller(2009a)}]{HallMiller2009a}Hall, P. and Miller, H. (2009a) Using generalized correlation to effect variable selection in very high dimensional problems. 
\newblock \emph{J. Comput. Graph. Statist.}, \textbf{18}, 533--550.

\bibitem[{Hall and Miller(2009b)}]{HallMiller2009b}Hall, P. and Miller, H. (2009b) Using the bootstrap to quantify the authority of an empirical ranking.
\newblock \emph{Ann. Statist.}, \textbf{37}, 3929--3959.



\bibitem[{Hall, Park and Samworth(2008)}]{HPS2008}Hall, P., Park, B. U. and Samworth, R. J. (2008) Choice of neighbor order in nearest-neighbor classification. 
\newblock \emph{Ann. Statist.}, \textbf{36}, 2135--2152.

\bibitem[{Hall and Pham(2010)}]{HallPham2010}Hall, P. and Pham, T. (2010). Optimal properties of centroid-based classifiers for very high-dimensional data. 
\newblock \emph{Ann. Statist.}, \textbf{38}, 1071--1093.

\bibitem[{Hall and Samworth(2005)}]{HallSamworth2005}Hall, P. and Samworth, R. J. (2005) Properties of bagged nearest neighbour classifiers. 
\newblock \emph{J. Roy. Statist. Soc. Ser. B}, \textbf{67}, 363--379.

\bibitem[{Hall, Titterington and Xue(2009a)}]{HTX2009a}Hall, P., Titterington, D. M. and Xue, J.-H. (2009a). Tilting methods for assessing the influence of components in a classifier. 
\newblock \emph{J. Roy. Statist. Soc. Ser. B}, \textbf{71}, 783--803.

\bibitem[{Hall, Titterington and Xue(2009b)}]{HTX2009b}Hall, P., Titterington, D. M. and Xue, J.-H. (2009b) Median-based classifiers for high-dimensional data. 
\newblock \emph{J. Amer. Statist. Assoc.}, \textbf{104}, 1597--1608.

\bibitem[{Hall, Xia and Xue(2013)}]{HXX2013}Hall, P., Xia, Y. and Xue, J.-H. (2013) Simple tiered classifiers. 
\newblock \emph{Biometrika}, \textbf{100}, 431--445.

\bibitem[{Hall and Xue(2010)}]{HallXue2010}Hall, P. and Xue, J.-H. (2010) Incorporating prior probabilities into high-dimensional classifiers. 
\newblock \emph{Biometrika}, \textbf{97}, 31--48.


\bibitem[{Li(1991)}]{Li1991}Li, K. C. (1991) Sliced inverse regression for dimension reduction.
\newblock \emph{J. Amer. Statist. Assoc.}, \textbf{86}, 316--327.

\bibitem[{Li, Zhong and Zhu(2012)}]{LZZ2012}Li, R., Zhong, W. and Zhu, L. (2012) Feature screening via distance correlation learning.
\newblock \emph{J. Amer. Statist. Assoc.}, \textbf{107}, 1129--1139.

\bibitem[{Marron, Todd and Ahn(2007)}]{MTA2007}Marron, J. S., Todd, M. J. and Ahn, J. (2007) Distance-weighted discrimination.
\newblock \emph{J. Amer. Statist. Assoc.}, \textbf{102}, 1267--1271.

\bibitem[{Meinshausen and B\"uhlmann(2010)}]{MeinshausenBuhlmann2010}Meinshausen, N. and B\"uhlmann, P. (2010) Stability selection.
\newblock \emph{J. Roy. Statist. Soc., Ser. B (with discussion)}, \textbf{72}, 417--473.

\bibitem[{Samworth(2005)}]{Samworth2005}Samworth, R. (2005) Small confidence sets for the mean of a spherically symmetric distribution.
\newblock \emph{J. Roy. Statist. Soc., Ser. B}, \textbf{67}, 343--361.

\bibitem[{Samworth(2012)}]{Samworth2012}Samworth, R. J. (2012) Optimal weighted nearest neighbour classifiers.
\newblock \emph{Ann. Statist.}, \textbf{40}, 2733--2763.

\bibitem[{Shah and Samworth(2013)}]{ShahSamworth2013}Shah, R. D. and Samworth, R. J. (2013) Variable selection with error control: Another look at Stability Selection.
\newblock \emph{J. Roy. Statist. Soc., Ser. B}, \textbf{75}, 55--80.

\end{thebibliography}
\end{document}